\documentclass[a4paper,twocolumn,10pt,accepted=2022-02-25]{quantumarticle}
\pdfoutput=1
\usepackage[utf8]{inputenc}
\usepackage[english]{babel}
\usepackage[T1]{fontenc}
\usepackage{amsmath}
\usepackage{hyperref}

\usepackage[numbers,sort&compress]{natbib}

\usepackage{amsmath,amssymb,amsfonts}
\usepackage{braket}
\usepackage{bm}
\usepackage{algorithm}
\usepackage{algpseudocode}

\newcommand{\ie}[0]{\textit{i.e. }}

\newcommand{\mf}[1]{\mathfrak{#1}}
\newcommand{\mc}[1]{\mathcal{#1}}
\newcommand{\mb}[1]{\mathbb{#1}}
\newcommand{\trm}[1]{\textrm{#1}}

\renewcommand{\vec}[1]{\bm{#1}}

\begin{document}

\title{Variational quantum amplitude estimation}

\author{Kirill Plekhanov}
\email{kirill.plekhanov@cambridgequantum.com}
\affiliation{Cambridge Quantum Computing Limited, SW1P 1BX London, United Kingdom}
\orcid{0000-0001-7790-9309}

\author{Matthias Rosenkranz}
\affiliation{Cambridge Quantum Computing Limited, SW1P 1BX London, United Kingdom}
\orcid{0000-0002-1605-9141}

\author{Mattia Fiorentini}
\affiliation{Cambridge Quantum Computing Limited, SW1P 1BX London, United Kingdom}
\orcid{0000-0001-5832-353X}

\author{Michael Lubasch}
\affiliation{Cambridge Quantum Computing Limited, SW1P 1BX London, United Kingdom}
\orcid{0000-0002-2636-9936}

\begin{abstract}
We propose to perform amplitude estimation with the help of constant-depth quantum circuits that variationally approximate states during amplitude amplification.
In the context of Monte Carlo (MC) integration, we numerically show that shallow circuits can accurately approximate many amplitude amplification steps.
We combine the variational approach with maximum likelihood amplitude estimation [Y.\ Suzuki \textit{et al.}, Quantum Inf.\ Process.\ \textbf{19}, 75 (2020)] in variational quantum amplitude estimation (VQAE).
VQAE typically has larger computational requirements than classical MC sampling.
To reduce the variational cost, we propose adaptive VQAE and numerically show in 6 to 12 qubit simulations that it can outperform classical MC sampling.
\end{abstract}

\maketitle

\section{Introduction}

Amplitude estimation~\cite{BrassardHoyer2000} is a powerful algorithm that can achieve a quadratic quantum speedup over classical Monte Carlo (MC) methods~\cite{Montanaro2015}.
It has a wide range of applications, e.g.\ in quantum chemistry~\cite{KnillEtAll2007, Kassal18681}, machine learning~\cite{WiebeEtAl2014, WiebeEtAl2014-2, KerenidisEtAl2018}, and finance~\cite{OrusEtAl2019, EgEtAl20, BoEtAl20} where it can help with tasks such as risk analysis~\cite{WoernerEtAl2019, BraunEtAl2021} and the pricing of financial derivatives~\cite{RebenstrotEtAl2018, StamatopoulosEtAl2019}.

The original amplitude estimation procedure~\cite{BrassardHoyer2000} has hardware requirements that are challenging for current quantum devices and, therefore, reducing these requirements is currently an active area of research.
Crucial breakthroughs were obtained in recent proposals which succeeded in replacing the hardware-intensive components of traditional amplitude estimation -- controlled multi-qubit gates and quantum Fourier transform -- by classical post-processing~\cite{SuzukiEtAl2020, tanaka2020amplitude, AaronsonRall2019, GrinkoEtAl2019}.
Alternatively, one can systematically reduce the circuit depth by interpolating between classical MC methods and amplitude estimation~\cite{Giurgica-TironEtAl2020}.
Additionally, classical pre-processing can replace costly quantum arithmetic~\cite{Herbert2021} and Bayesian inference can be used to boost the algorithmic efficiency in the presence of device errors~\cite{WangKohEtAl21, KatabarwaEtAl21}.

\begin{figure}[!t]
\centering
\includegraphics[width=.95\columnwidth]{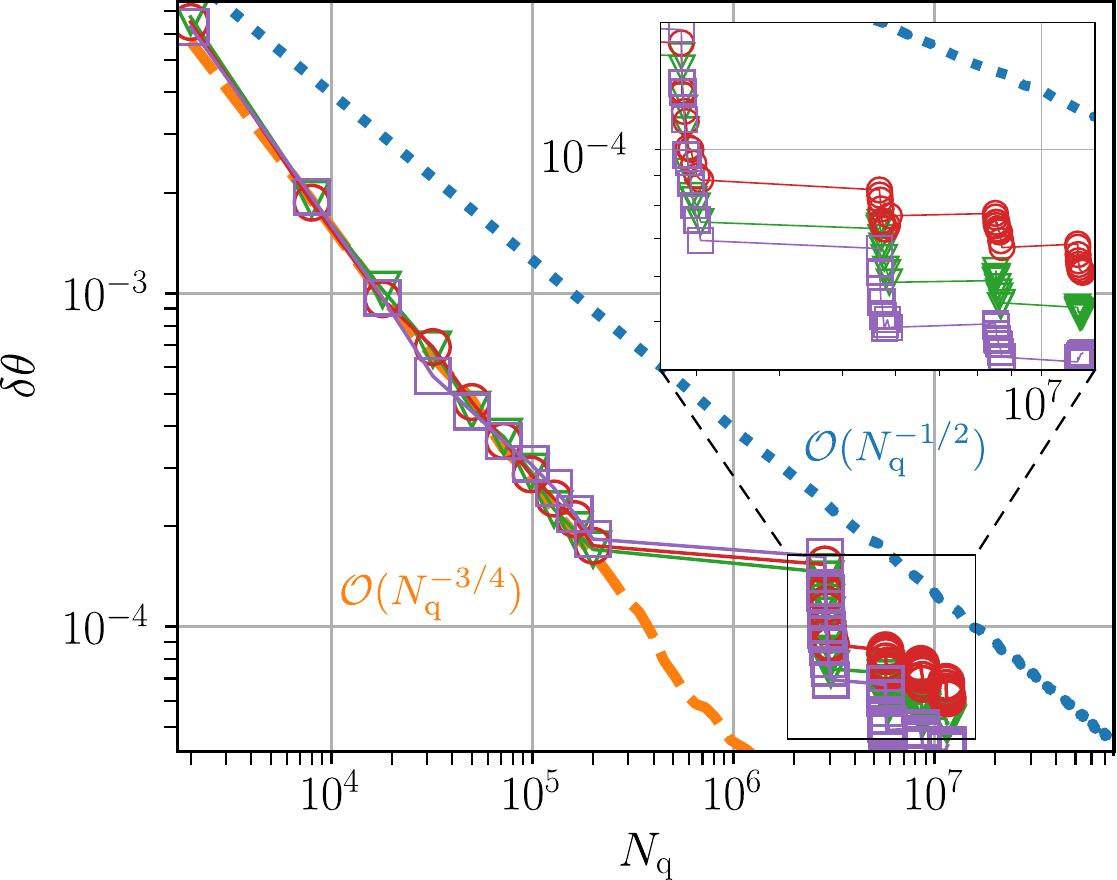}
\caption{\label{fig:adaptive}
Amplitude estimation error $\delta \theta$ as a function of the computational cost, i.e.\ the number of queries $N_{\trm{q}}$.
We compare adaptive VQAE (symbols with lines guide to the eye) to MLAE (dashed orange) and classical MC sampling (dotted blue).
We calculate the rescaled mean value of a shifted Cauchy-Lorentz (red circles), Gaussian (green triangles), and log-normal (purple squares) probability distribution.
In VQAE, the first $10$ amplitude estimates are computed via MLAE, then one step of the variational optimization is performed, which is followed by the next iteration of $10$ MLAE steps.
This procedure of variational approximation followed by MLAE is repeated three more times, resulting in a final error $\delta \theta \approx 6 \cdot 10^{-5}$.
We see that, for this error, $N_{\trm{q}}$ is up to an order of magnitude smaller in VQAE than in classical MC sampling.
Throughout this calculation, VQAE's circuit depth is the depth of the initial state plus at most $10$ times the depth of the query operator, whereas MLAE's circuits have the depth of the initial state plus, at the end, $50$ times the depth of the query operator.
}
\end{figure}

In this article, we address the question whether the quantum computational requirements for amplitude estimation can be further decreased by making use of variational quantum algorithms~\cite{BeEtal19, CeEtAl20, BhEtAl21}.
We present variational quantum amplitude estimation (VQAE) in which the depth of the entire quantum circuit is always kept below a desired maximum value by means of variational optimization.
VQAE is based on maximum likelihood amplitude estimation (MLAE)~\cite{SuzukiEtAl2020}.
We present a na\"{i}ve and an adaptive VQAE algorithm.
Adaptive VQAE rescales the amplitude to reduce the cost of the variational optimization.
The advantage of VQAE over MLAE is that the maximum circuit depth of VQAE is independent of the total number of MLAE steps, whereas in MLAE this depth grows linearly with the number of MLAE steps.
The advantage of VQAE over classical MC sampling is that VQAE can have a lower computational cost.
Figure~\ref{fig:adaptive} shows that, for the problems considered here, VQAE outperforms classical MC sampling and additionally keeps the overall circuit depth below a fixed value.

This article is organized as follows.
Firstly, in Section~\ref{sec:problem}, we define the problems considered here and explain the original quantum algorithm for amplitude estimation as well as the classical MC approach. 
Then, in Section~\ref{sec:variational} we present our variational methods, study variational errors of constant-depth quantum circuits, and develop na\"{i}ve and adaptive VQAE.
We conclude this article and discuss potential next steps in Section~\ref{sec:Discussion}.

\section{Background}
\label{sec:problem}

In this section, we first define the problem that we are interested in.
Next, we explain quantum amplitude estimation and classical MC sampling.

\subsection{Problem definition}

Throughout this article, we focus on the calculation of expectation values
\begin{align}\label{eq:expectation_value}
 \mb{E}_p[f] = \sum_x p(x) f(x)
\end{align}
where the sum runs over $2^{n}$ equidistant values of $x \in [0, 1)$, $p(x)$ represents a probability distribution and $f(x)$ a real-valued function.
Here $n$ is the qubit count of the wave function that encodes $p(x)$ and $f(x)$ in its amplitudes.
We consider three probability distributions:
a Gaussian
\begin{align}\label{eq:problem_p_1}
 p_{\text{G}}(x) = \frac{1}{\mc{N}_{\text{G}}} \exp \left( -\frac{(x  - \mu)^{2}}{2 \sigma^{2}} \right) ,
\end{align}
Cauchy-Lorentz
\begin{align}\label{eq:problem_p_2}
 p_{\text{C-L}}(x) = \frac{1}{\mc{N}_{\text{C-L}}} \frac{\sigma}{(x - \mu)^2 + \sigma^2} ,
\end{align}
and log-normal distribution
\begin{align}\label{eq:problem_p_3}
 p_{\text{l-n}}(x) = \frac{1}{\mc{N}_{\text{l-n}}(c_{0} + c_{1} x)} \exp \left( -\frac{(\ln(c_{0} + c_{1} x) - \mu)^{2}}{2 \sigma^{2}} \right) .
\end{align}
The normalization constants $\mc{N}_{\text{G}}$, $\mc{N}_{\text{C-L}}$ and $\mc{N}_{\text{l-n}}$ are chosen so that $\sum_x p(x) = 1$.

We choose the following parameters for our analysis.
We fix the total number of qubits encoding $f(x)$ and $p(x)$ to $n = 5$.
In our calculations with the Gaussian and Cauchy-Lorentz distribution, we use $\mu = 0.5$ and $\sigma = 0.1$.
In our calculations with the log-normal distribution, we use $c_{0} = 0$, $c_{1} = 10$, $\mu = 1.5$, and $\sigma = 0.2$.
For the function $f(x)$, we use
\begin{align}\label{eq:problem_f}
 f(x) = C x
\end{align}
with some $C > 0$.
For this choice of parameters, the expectation value~\eqref{eq:expectation_value} is approximately $\mb{E}_p[f] \approx 0.5\ C$ for all distributions.

\subsection{Quantum amplitude estimation}

Let us present a way to encode the solution to~\eqref{eq:expectation_value} on a quantum computer.
We assume $f(x)$ and $p(x)$ are functions that map $[0, 1)$ to $[0, 1]$.
We consider real numbers $x \in [0, 1)$ that satisfy
\begin{align}
 x = \sum_i \frac{x_i}{2^{n-i}},\quad x_i \in \{ 0, 1 \}
\end{align}
and that we identify with $n$-bit strings $\lbrace x_i,\ i = 1,\ 2,\ \dots,\ n \rbrace$.
Each bit string shall correspond to a quantum state $\ket{x} = \ket{x_{1},\ x_{2},\ \ldots,\ x_{n}}$ in the computational basis of a $n$-qubit register.
Additionally we have a quantum circuit $\mc{A}$ that acts on a register of $n+1$ qubits and produces a state $\ket{\chi_0}_{n+1} = \mc{A} \ket{0}_{n+1}$ such that
\begin{align}\label{eq:chi_0}
 \ket{\chi_0}_{n+1} = \sqrt{1 - a} \ket{\psi_{\trm{bad}}}_n \ket{0} + \sqrt{a} \ket{\psi_{\trm{good}}}_n \ket{1}.
\end{align}
Here $\ket{\psi_{\trm{bad}}}_n$ and $\ket{\psi_{\trm{good}}}_n$ are two normalized quantum states of a $n$-qubit register which is connected to one additional ancilla qubit.
We define the good state
\begin{align}\label{eq:psi_good}
 \ket{\psi_\trm{good}}_n = \frac{1}{\sqrt{a}} \sum_x \sqrt{p(x) f(x)} \ket{x}_n
\end{align}
so that $a = \mb{E}_{p}[f]$ of Eq.~\eqref{eq:expectation_value} coincides with the probability of measuring the ancilla qubit in the state $\ket{1}$.

To determine $a$, the amplitude estimation algorithm uses the Grover operator $\mc{Q} = - \mc{R}_{\chi} \mc{R}_{\trm{good}}$~\cite{BrassardHoyer1997, Grover1998, BrassardHoyer2000} where
\begin{align}\label{eq:reflections}
&
\mc{R}_{\chi} = \mathbb{I} - 2 \ket{\chi_0}_{n+1}\bra{\chi_0}_{n+1} = \mathbb{I} - 2 \mc{A} \ket{0}_{n+1} \bra{0}_{n+1} \mc{A}^{\dag} \notag \\
&
\mc{R}_{\trm{good}} = \mathbb{I} - 2 \ket{\psi_{\trm{good}}}_n \ket{1} \bra{\psi_{\trm{good}}}_n \bra{1}
\end{align}
are reflections in a two-dimensional subspace $\mc{H}_{\chi}$ spanned by states $\ket{\psi_{\trm{bad}}}_n \ket{0}$ and $\ket{\psi_{\trm{good}}}_n \ket{1}$.
We define $a = \sin^2(\theta)$ and explicitly write out the action of $Q$:
\begin{align}\label{eq:q_to_the_power_m}
\ket{\chi_m}_{n+1} = Q^{m} \ket{\chi_0}_{n+1} = & \cos [(2m+1) \theta] \ket{\psi_{\trm{bad}}}_n \ket{0} \notag \\ + & \sin [(2m+1) \theta] \ket{\psi_{\trm{good}}}_n \ket{1}.
\end{align}
Therefore, the subspace $\mc{H}_{\chi}$ is stable under the action of $\mc{Q}$ and the only effect of $\mc{Q}$ is to rotate by an angle of $2 \theta$.
The original amplitude estimation algorithm then uses quantum phase estimation to find the eigenvalues of $\mc{Q}$ equal to $\exp(\pm 2 \mathrm{i} \theta)$ and provides an estimate of $a$ with an error
\begin{align}
 \epsilon \leq
 2 \pi \frac{\sqrt{a (1-a)}}{2 N_{\trm{q}}} +
 \frac{\pi^2}{4 N_{\trm{q}}^{2}},
\end{align}
with a probability of at least $8 / \pi^2$~\cite{BrassardHoyer2000} where $N_{\trm{q}}$ ($2 N_{\trm{q}}$) is the total number of times the operator $\mc{A}$ ($\mc{Q}$) has to be applied.

Traditional amplitude estimation has high requirements on quantum hardware because it uses quantum phase estimation.
This algorithm needs the quantum Fourier transform and multiple controlled $\mc{Q}^{m}$ operations where $\lbrace m = 1,\ 2,\ 4,\ \dots,\ 2^{M} \rbrace$.
The depth of the corresponding quantum circuit is mostly determined by the depth of the last controlled $\mc{Q}^{m}$ operator for which $m = 2^{M}$.
In general, the total circuit depth scales like the total number of queries $\mc{O}(N_{\trm{q}}) \sim \mc{O}(1 / \epsilon)$ inversely proportional to the desired error $\epsilon$.

To avoid these deep quantum circuits, several recent articles propose new ways to carry out amplitude estimation circumventing quantum phase estimation~\cite{SuzukiEtAl2020, AaronsonRall2019, GrinkoEtAl2019} and circuits of depth $\mc{O}(1 / \epsilon)$~\cite{Giurgica-TironEtAl2020}.
One proposal is MLAE~\cite{SuzukiEtAl2020} in which one combines measurements of the states $\ket{\chi_m}_{n+1}$ with a maximum likelihood estimation of $a$.
For an exponential schedule $\lbrace m = 1,\ 2,\ 4,\ \dots,\ 2^{M} \rbrace$, this algorithm has the query cost $N_{\trm{q}} \sim \mc{O}(1 / \epsilon)$.
A linear schedule $\lbrace m = 1,\ 2,\ 3,\ \dots,\ M \rbrace$ increases the query cost to $N_{\trm{q}} \sim \mc{O}(\epsilon^{-4/3})$.
Note that in this case $N_{\trm{q}}$ scales quadratically with the maximum circuit depth $M$.
Following the same idea of reducing the hardware requirements, the authors of Ref.~\cite{Giurgica-TironEtAl2020} present two algorithms with computational cost $N_{\trm{q}} \sim \mc{O}(1 / \epsilon^{\beta+1})$ for quantum circuits of reduced depth $\mc{O}(1 / \epsilon^{1 - \beta})$.
These algorithms are controlled by an external parameter $\beta$ which allows one to interpolate between the quantum regime at $\beta = 0$ and the classical MC regime at $\beta = 1$.

\subsection{Classical MC sampling}

We perform classical MC sampling in the following way.
We sample from the state $\ket{\chi_0}_{n+1}$ of Eq.~\eqref{eq:chi_0} and measure the ancilla qubit.
We compute $a$ as the relative frequency of measuring the ancilla qubit in the state $\ket{1}$.
This calculation of $a$ has the error $\epsilon = \sqrt{a (1 - a) / N_{\trm{q}}}$ so that the total number of queries required for a certain error $\epsilon$ is $N_{\trm{q}} \sim \mc{O}(1 / \epsilon^2)$~\cite{PrEtAl92}.

Comparing this query cost with the previous ones, we find that traditional amplitude estimation as well as MLAE with exponential schedule achieve a quadratic quantum speedup over classical MC sampling.
Both MLAE with linear schedule and the algorithms in~\cite{Giurgica-TironEtAl2020} obtain a reduced quantum speedup.

Note that, throughout this article, the query complexity is defined in terms of $\mc{A}$ operators, with two applications of $\mc{A}$ required per application of $\mc{Q}$, see Eq.~\eqref{eq:reflections}.
Also, the depth of quantum circuits is measured in units of $\mc{A}$, so that the depth of $\ket{\chi_0}_{n+1}$ is equal to one and the depth of $\mc{Q}$ is equal to two.
Additionally, in the following we assume that $\mc{A}$ and $\mc{Q}$ are given, i.e.\ we do not address questions e.g.\ relating to their efficient quantum circuit respresentation.

\section{Variational algorithms}
\label{sec:variational}

Here we present our variational algorithms.
We first explain the general VQAE formalism, then our na\"{i}ve implementation, and finally the adaptive VQAE approach.

\subsection{General formalism}

The VQAE algorithm is based on the maximum likelihood framework of Ref.~\cite{SuzukiEtAl2020} with linearly incremental sequence $\lbrace m = 1,\ 2,\ 3,\ \dots,\ M \rbrace$.
In this framework, the depth of the quantum circuit implementing the state $\ket{\chi_m}_{n+1} = \mc{Q}^{m} \ket{\chi_0}_{n+1}$ scales with $m$ as $2 m + 1$.
To prevent the circuit depth from increasing indefinitely, we add to this framework a variational step during which states $\ket{\chi_m}_{n+1}$ are periodically approximated by a variational quantum state of depth one.
We note that this strategy will not always work and the corresponding approximation can have a large error.
The variational approach, however, allows us to compute the approximation error so that we can identify when the strategy works.
We perform the variational approximation every $k$-th power of $\mc{Q}$, with $0 < k < M$.
For all the other iterations, we simply apply the corresponding power of $\mc{Q}$ to the variational state.
This results in Algorithm~\ref{algorithm_1}.

\begin{algorithm}[H]
\caption{Variational quantum amplitude estimation}
\label{algorithm_1}
\begin{algorithmic}
\Require functions $f$ and $p$, integer $k$
\State - Use $f$ and $p$ to encode
$\ket{\phi_{i=0}}_{n+1} = \ket{\chi_0}_{n+1}$
\State and $\mc{Q} = -\mc{R}_{\chi} \mc{R}_{\trm{good}}$ according to Eqs.~\eqref{eq:chi_0} and~\eqref{eq:reflections}
\For{$0 \leq m \leq M$}
    \State - Set $i = \lfloor m / k \rfloor, \ j = m \% k$
    \State \Comment{sampling}
    \State - Sample the circuit $\mc{Q}^j \ket{\phi_i}_{n+1}$ and collect $h$ samples.
    \State - Save the number of times the ancilla qubit is $\ket{1}$ in
    \State a variable $h_m$
    \State \Comment{end sampling}
    \State \Comment{variational approximation}
    \If{$j = k-1$}
    \State - Perform the variational approximation
    \begin{equation*}
        \ket{\phi_{i+1}}_{n+1} \approx \mc{Q}^{k} \ket{\phi_{i}}_{n+1}
    \end{equation*}
    \EndIf
    \State \Comment{end variational approximation}
\EndFor
\State - Use $\lbrace h_m \rbrace$ to carry out the maximum likelihood estimation
\end{algorithmic}
\end{algorithm}

\noindent Here $\lfloor \cdot \rfloor$ denotes the floor function and $\%$ the modulo operation.
The resulting approximation of $\ket{\chi_m}_{n+1}$ corresponds to the state $\mc{Q}^{j} \ket{\phi_i}_{n+1}$ with $i = \lfloor m / k \rfloor$ and $j = m \% k$.
The depth of this approximation reaches the minimum of one when $j = 0$ and the maximum of $2k - 1$ when $j = k-1$.

\begin{figure*}
\centering
\includegraphics[width=141.864mm]{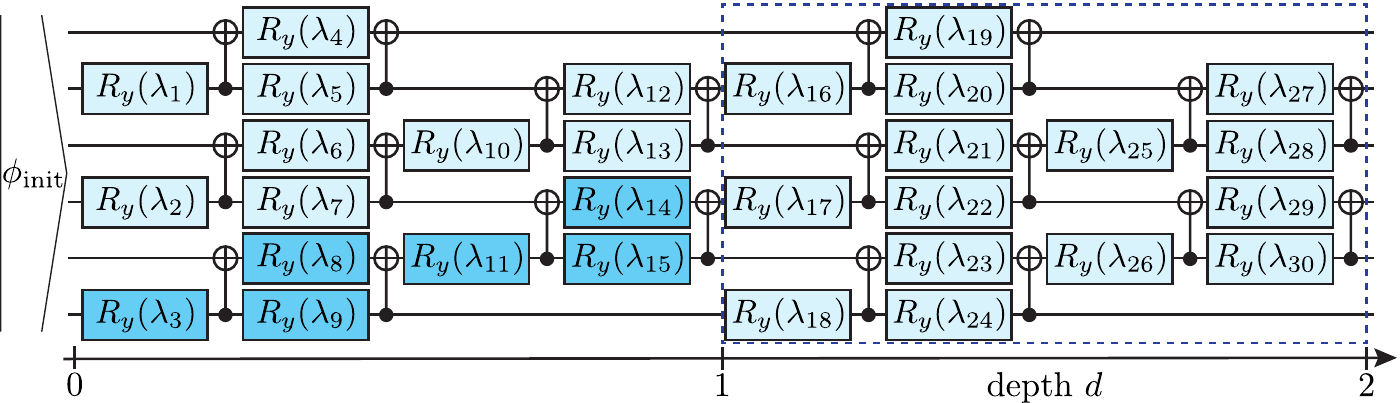}
\caption{\label{fig:pqc}
Variational PQC ansatz of depth $d$ for $6$ qubits.
The variational parameters $\lambda_{j}$ are in single-qubit rotation gates $R_{y}(\lambda_{j}) = \exp(-\mathrm{i} \lambda_{j} \sigma_{y} / 2)$ where $\sigma_{y}$ denotes the $y$ Pauli matrix.
In na\"{i}ve VQAE, $\ket{\phi_{\trm{init}}}_{n+1} = \ket{0}_{n+1}$ and the circuit structure inside the dashed box is repeated $d$ times.
In adaptive VQAE, $\ket{\phi_{\trm{init}}}_{n+1} = \ket{\chi_0}_{n+1}$ and only the dark blue gates with adjacent CNOTs form our variational ansatz.}
\end{figure*}

The maximum likelihood post-processing~\cite{SuzukiEtAl2020} consists in maximizing the likelihood function 
$L (\lbrace h_m \rbrace, x) = \prod_m L_m (h_m, x)$ with
\begin{align}
 L_m (h_m, x) = [\sin^2((2m+1) x)]^{h_m} [\cos^2((2m+1) x)]^{h - h_m} ,
\end{align}
so that the estimate of the phase $\theta$ becomes
\begin{align}
 \hat{\theta} = \arg \max_x \Big( \ln L (\lbrace h_m \rbrace, x) \Big) .
\end{align}
Our implementations of the maximum likelihood estimation use $h = 2 \times 10^3$ samples.
The minimization of $L (\lbrace h_m \rbrace, x)$ is accomplished by means of a brute-force search algorithm that uses $5 \times 10^3$ grid points.

We variationally approximate states $\mc{Q}^k \ket{\phi_i}_{n+1}$ by minimizing $||\ket{\phi_{\trm{var}} (\vec{\lambda})}_{n+1} - \mc{Q}^{k} \ket{\phi_i}_{n+1}||^{2}$ which is equivalent to maximizing the objective function
\begin{align}\label{eq:objective_function}
 \mc{F}(\vec{\lambda}) = \mf{Re} \Big( \bra{\phi_{\trm{var}} (\vec{\lambda})}_{n+1} \mc{Q}^{k} \ket{\phi_i}_{n+1} \Big)
\end{align}
with respect to the variational parameters $\vec{\lambda}$.
The optimal solution can be formally written as 
\begin{align}
 \ket{\phi_{i+1}}_{n+1} = \ket{\phi_{\trm{var}}(\tilde{\vec{\lambda}})}, \quad
 \tilde{\vec{\lambda}} = \arg\max\limits_{\vec{\lambda}} \mc{F}(\vec{\lambda}) .
\end{align}
We notice that the depth of the quantum circuit required to compute $\mc{F}(\vec{\lambda})$ is the largest circuit depth used by the algorithm.
This quantum circuit is composed of the parts encoding $\ket{\phi_{\trm{var}}(\vec{\lambda})}_{n+1}$ and $\ket{\phi_i}_{n+1}$, each having depth one, and an operator $\mc{Q}^k$ of depth $2 k$, resulting in a total depth of $2k + 2$.
In general, the variational quantum state $\ket{\phi_{\trm{var}} (\vec{\lambda})}_{n+1}$ is a parameterized quantum circuit (PQC)
\begin{align}\label{eq:parametrized_circuit}
 \ket{\phi_{\trm{var}}(\vec{\lambda})}_{n+1} = \mc{U}_{\trm{var}}(\vec{\lambda})
 \ket{\phi_{\trm{init}}}_{n+1} = \prod_j e^{-\mathrm{i} \lambda_j G_{j}} \ket{\phi_{\trm{init}}}_{n+1} ,
\end{align}
where $G_j$ are Hermitian operators acting on the $(n+1)$-qubit register and $\ket{\phi_{\trm{init}}}_{n+1}$ is some initial state.
For our purposes, we are interested in hardware-efficient quantum circuits that produce real-valued quantum states.
We use the PQC shown in Fig.~\ref{fig:pqc} that is composed of $d$ layers with $15$ parameterized single-qubit rotation gates and $10$ CNOT gates per layer.

One single variational update of a PQC consists of $n_{\trm{s}}$ sweeps over all circuit parameters, during which all parameters are updated simultaneously.
To perform the optimization, it is convenient to introduce a coordinate-wise version of Eq.~\eqref{eq:objective_function} for the $j$-th parameter 
\begin{align}
 f_j(x) = \mathcal{F}(\lambda_1,\ \lambda_2,\ \dots,\ \lambda_{j-1},\ x,\ \lambda_{j+1},\ \dots).
\end{align}
The optimization of the parameterized state in Eq.~\eqref{eq:parametrized_circuit} can then be performed via a particle swarm approach~\cite{KennedyEberhart1995, BonyadiRezaZbigniew2017}, the coordinate-wise update~\cite{VidalEtAl2018, PaEtAl19, NakanishiEtAl2020, OstaszewskiEtAl2019, BenedettiEtAl2020}, or gradient based methods with the parameter-shift rule~\cite{hazan2015beyond, suzuki2021normalized, LiEtAl2017, MitaraiEtAl2018, SchuldEtAl2019, BanchiEtAl2020}
\begin{align}\label{eq:gradient_descent}
 \frac{\trm{d} f_j(\lambda_j)}{\trm{d} \lambda_j} & = f_j(\lambda_j + \pi / 4) - f_j(\lambda_j - \pi / 4) .
\end{align}
We obtained the best results using the gradient based approach with the Adam optimizer~\cite{KingmaBa2014}.
Therefore this technique is being used throughout this article for the computation of all results.
Each gradient calculation requires two evaluations of the coordinate-wise objective function $f_j(\lambda_j \pm \pi / 4)$.
On a quantum computer, $f_j$ can be determined via the Hadamard test~\cite{AharonovEtAl2005}.
In our numerical simulations, we emulate the measurement of the Hadamard circuit by first evaluating the exact value of $f_j$ and then sampling it using a binomial distribution with the probability $(1 + f_{j}) / 2$ and $n_{f}$ independent Bernoulli trials~\cite{PrEtAl92}.

The variational approximation step significantly affects the total number of queries $N_{\trm{q}}$ used by VQAE.
In MLAE with a linearly incremental sequence, the total number of queries is equal to
\begin{align}\label{eq:M_scaling_maximum_likelihood}
 N_{\trm{q}} = \sum_{m = 1}^{M} h (2 m + 1) = h M (M + 2) ,
\end{align}
where $2 m + 1$ is the depth of the quantum circuit encoding $\ket{\chi_m}_{n+1} = \mc{Q}^{m} \ket{\chi_0}_{n+1}$.
In VQAE, the total number of queries is composed of two separate contributions.
The first one accounts for the sampling of the quantum circuits $\mc{Q}^{j} \ket{\phi_i}_{n+1}$ and we denote it by $N_{\trm{samp}}$.
The corresponding section in Algorithm~\ref{algorithm_1} is labelled by ``sampling''.
The second contribution corresponds to the variational approximation cost, which we denote by $N_{\trm{var}}$.
It is associated with the section in Algorithm~\ref{algorithm_1} labelled by ``variational approximation''.
We assume that the number of queries required per variational approximation is independent of the iteration number $m$ and changes only as a function of the desired variational error as well as the depth of the circuit for the objective function.
We denote the cost of a single variational update as $N_{\trm{var/1}} (2 k + 2)$, where $(2 k + 2)$ is the depth of the objective function $\mc{F}(\vec{\lambda})$ and $N_{\trm{var/1}}$ is the number of quantum circuits per variational update that need to be run by the algorithm.
As a result, the total number of variational queries becomes
\begin{align}
 \label{eq:nvarp1}
 N_{\trm{var}} =
 N_{\trm{var/1}} (2k + 2) \lfloor M / k \rfloor
 \sim
 \mc{O}(k \lfloor M / k \rfloor ) ,
\end{align}
where $\lfloor M / k \rfloor$ is the total number of variational updates required to approximate $\mc{Q}^M$.
The number of sampling queries is equal to
\begin{align}
 N_{\trm{samp}}
 &=
 \lfloor M / k \rfloor \sum_{j = 0}^{k-1} h (2 j + 1) +
 \sum_{j = 0}^{M \% k} h (2 j + 1)
 \notag \\
 &=
 h k (k + 2) \lfloor M / k \rfloor + h (M \% k ) (M \% k + 2),
\end{align}
where the last term accounts for the situation when $k$ is not a divisor of $M$.
In the limit $M \gg k$, $N_{\trm{var}} \sim \mc{O}(M)$ and $N_{\trm{samp}} \sim \mc{O}(kM)$. Note that both contributions scale like $\mc{O}(M)$ which is quadratically better than the scaling $\mc{O}(M^{2})$ of MLAE in Eq.~\eqref{eq:M_scaling_maximum_likelihood}.

\subsection{Na\"{i}ve VQAE}
\label{sec:naive}

\begin{figure*}
\centering
\includegraphics[width=1.94\columnwidth]{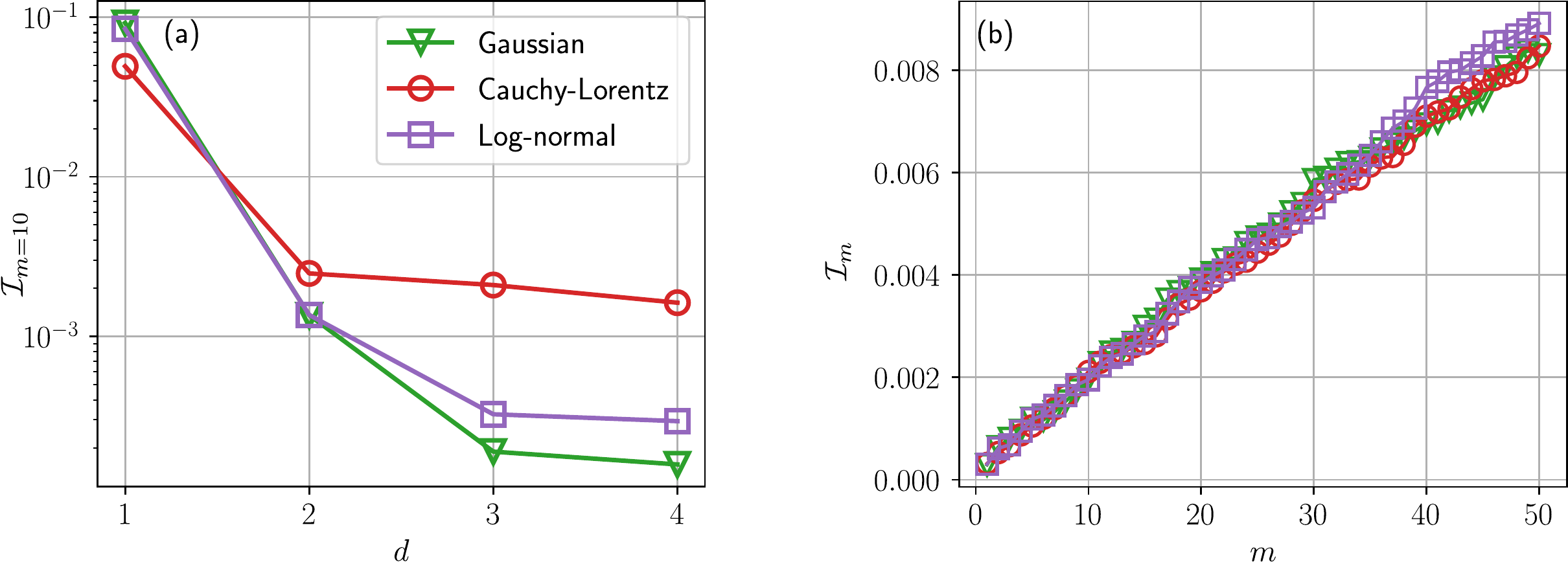}
\caption{\label{fig:infidelity}
Infidelity $\mc{I}_m$ in Eq.~\eqref{eq:infidelity} for the probability distributions of Eqs.~\eqref{eq:problem_p_1}-\eqref{eq:problem_p_3} (see legend) shown (a) as a function of $d$ for $m = 10$ and (b) as a function of $m$ for $d = 4$.
We see that the infidelity decreases with increasing $d$, due to the corresponding increase of the expressive power of the PQC.
The infidelity increases linearly as a function of $m$ slowly with a slope of $\approx 0.00017$ that is approximately the same for the three distributions.
These results are obtained via na\"{i}ve VQAE with $k = 1$, $M = 50$, and $n_{\trm{s}} = 1000$, using the numerically exact gradient without sampling and Adam with the initial learning rate $\beta = 0.1$.
We consider $100$ randomly initialized PQC and (a) shows one example calculation and (b) the average over all $100$ calculations.}
\end{figure*}

In our na\"{i}ve implementation of the VQAE algorithm, the initial state of the PQC in Eq.~\eqref{eq:parametrized_circuit} and Fig.~\ref{fig:pqc} is $\ket{\phi_{\trm{init}}}_{n+1} = \ket{0}_{n+1}$.

Let us first explore the expressive power of the corresponding variational state $\ket{\phi_{\trm{var}}(\vec{\lambda})}$.
To this end, we perform amplitude amplifications followed by variational approximations of the resulting state with $k = 1$ and $M = 50$.
To evaluate the quality of the variational approximation, we calculate the infidelity
\begin{align}\label{eq:infidelity}
 \mc{I}_m = 1 - \bra{\chi_m}_{n+1} Q^{j} \ket{\phi_{i}}_{n+1},\ m = i \cdot k + j,
\end{align}
where for $k = 1$ we have $j = 0$ and $i = m$. Figure~\ref{fig:infidelity}(a) shows the results of such calculations performed for different depths $d$ of the PQC for $m = 10$ and Fig.~\ref{fig:infidelity}(b) shows the infidelity as a function of $m$ for $d = 4$.
We observe that the accuracy of the variational ansatz increases with the depth and saturates at $d \approx 4$.
The infidelity increases linearly with $m$.
This behaviour is seen for all probability distributions considered.

\begin{figure}
\centering
\includegraphics[width=.99\columnwidth]{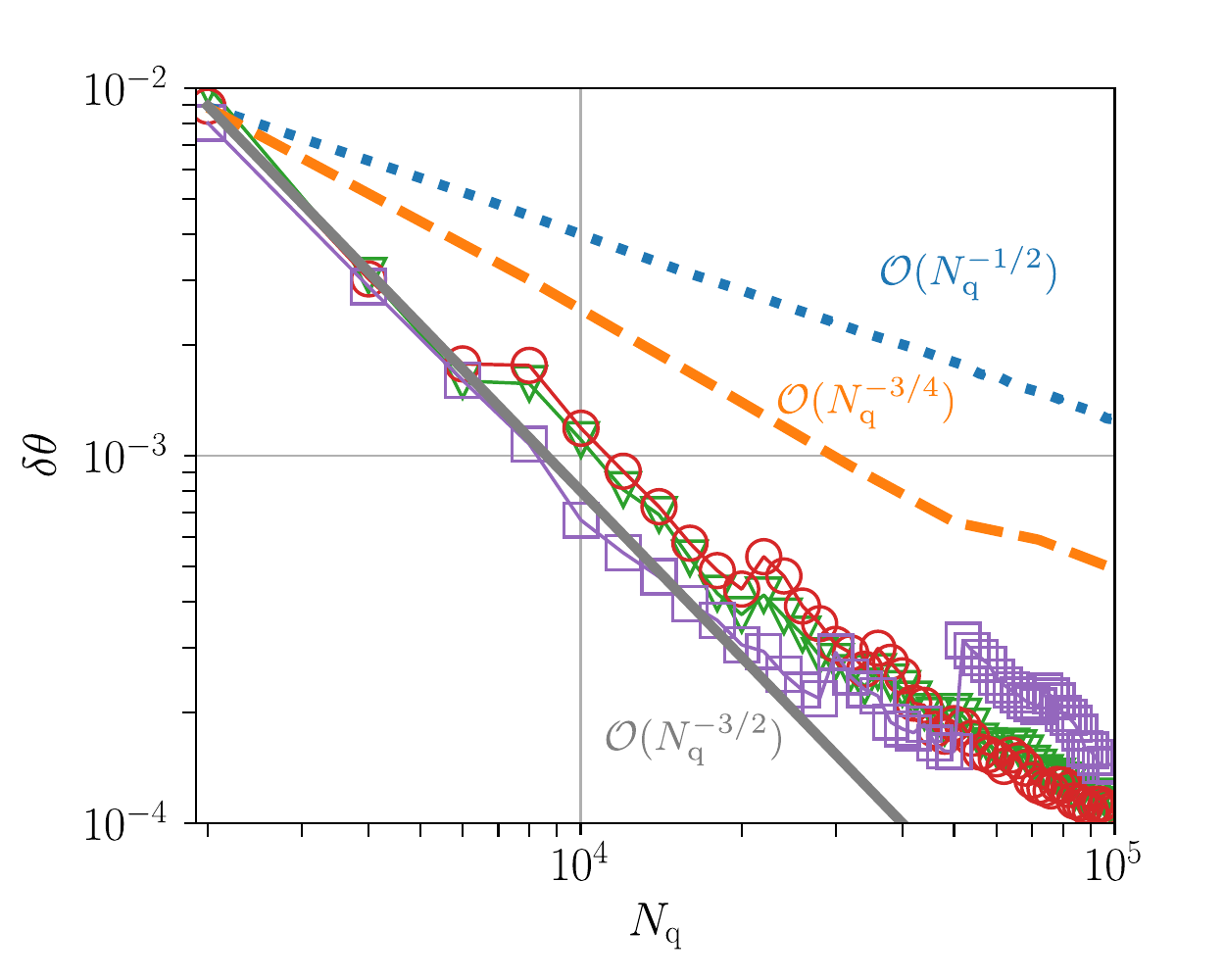}
\caption{\label{fig:naive}
Amplitude estimation error $\delta \theta$ as a function of the number of queries $N_{\trm{q}}$ obtained using na\"{i}ve VQAE with $k = 1$ under the assumption of zero variational cost $N_{\trm{var}/1} = 0$.
We observe that for small $M$, the error follows the ideal VQAE scaling $\delta \theta \sim \mc{O}(N_{\trm{q}}^{-3/2})$ (solid gray line).
For larger values of $M$, the scaling changes to the MC scaling $\delta \theta \sim \mc{O}(N_{\trm{q}}^{-1/2})$ (dotted blue line).
The result is also compared to the MLAE scaling $\delta \theta \sim \mc{O}(N_{\trm{q}}^{-3/4})$ (dashed orange line).
The legend as well as the simulation parameters are the same as in Fig.~\ref{fig:infidelity}.
}
\end{figure}

Next, we present the amplitude estimation results of na\"{i}ve VQAE.
Figure~\ref{fig:naive} shows the convergence of $\delta \theta$ as a function of $N_{\trm{q}}$, under the assumption that $N_{\trm{var}/1} = 0$ and $k = 1$. The resulting error is compared with the one of classical MC sampling which scales like $\delta \theta \sim \mc{O}(N_{\trm{q}}^{-1/2})$ and the one of MLAE which scales like $\delta \theta \sim \mc{O}(N_{\trm{q}}^{-3/4})$.
Interestingly, we find that the convergence of $\delta \theta$ changes as a function of $M$.
For small values of $M$, it follows the ideal VQAE scaling $\delta \theta \sim \mc{O}(N_{\trm{q}}^{-3/2})$ as if the variational approximation is performed without error.
We emphasize that this scaling is cubically better than the one of classical MC sampling.
The second convergence regime is observed for larger values of $M$.
In this regime, the error follows the MC scaling with $\delta \theta \sim \mc{O}(N_{\trm{q}}^{-1/2})$.
To understand this behaviour, we first notice that the MLAE error decreases with $M$, while the variational error increases instead.
In the regime when the MLAE error is larger than the variational error, the scaling of $\delta \theta$ is the best achievable MLAE scaling $\delta \theta \sim \mc{O}(N_{\trm{q}}^{-3/2})$. When the MLAE error is smaller than the variational error, the convergence of $\delta \theta$ is dominated by the latter.
The accumulation of the variational error can be modelled via a random process, in which each variational approximation results in a random error of zero mean and some variance $\sigma^2$. After $M$ steps of the algorithm, $\lfloor M / k \rfloor = M$ (as $k = 1$ here) variational approximations were performed resulting in a final error of variance $M \sigma^2$.
Hence, an ideal MLAE estimation of the angle $\theta$ will produce a relative error $\sqrt{M} \sigma / [(2M+1) \theta]$ scaling as $\mc{O}(M^{-1/2})$.
In our simulations, we find that the transition from the first regime -- where $\delta \theta \sim \mc{O}(N_{\trm{q}}^{-3/2})$ -- to the second regime -- where $\delta \theta \sim \mc{O}(M^{-1/2})$ -- occurs at $M \sim 20$.

Finally, we take into account the cost of the variational approximation, to obtain a more complete assessment of the algorithmic performance of na\"{i}ve VQAE.
To estimate the cost of a single variational update, we write down the number of circuits needed to be run for each variational update as $N_{\trm{var}/1} = 2 n_{f} n_{\trm{s}} n_{\trm{p}}$ where $n_{\trm{p}}$ is the number of parameters of a PQC, $n_{\trm{s}}$ is the total number of sweeps through all the parameters of the PQC, and $n_{f}$ is the number of Bernoulli trials per evaluation of the objective function.
The factor $2$ comes from the fact that two evaluations of the objective function are required for each evaluation of the gradient in Eq.~\eqref{eq:gradient_descent}.
For the PQC in Fig.~\ref{fig:pqc} with $d = 4$, the number of parameters is $n_{\trm{p}} = 60$.
Additionally, we choose $n_{f} \sim n_{\trm{s}} \sim 100$ so that $N_{\trm{var}/1} \sim 1.2 \times 10^6$ and $N_{\trm{var}} \sim 4.8 \times 10^{6} M$.
This large variational cost is the dominant part in the calculation of the total number of queries $N_{\trm{q}}$.
Ultimately, it leads to a performance of na\"{i}ve VQAE that is worse than the one of classical MC sampling.
Reducing any of $n_{f}$, $n_{\trm{s}}$, or $n_{\trm{p}}$ decreases the variational cost but also increases the variational error which then leads to a worse final amplitude estimation error.

\subsection{Adaptive VQAE}
\label{sec:adaptive}

To reduce the variational cost of VQAE, in the following we present the adaptive VQAE algorithm.
In this algorithm, the function $f$ is rescaled such that the $k$-th power of the Grover operator is close to the identity and only then the variational optimization is carried out.
Then the PQC ansatz needs to be just slightly different from the initial state $|\phi_{\trm{init}}\rangle$ and the optimization needs fewer samples than na\"{i}ve VQAE.

To introduce the adaptive VQAE algorithm, we first note that the amplitude $a = \mb{E}_{p}[f]$ -- see Eq.~\eqref{eq:expectation_value} -- is linear in $f$, meaning that rescaling the function $f$ with a proportionality constant $r$ also rescales the amplitude $a$:
\begin{align}\label{eq:rescaling}
 a' = \mb{E}_{p}[f'] = \mb{E}_{p}[r f] = r a .
\end{align}
The rescaled function $f'$ can then be used to encode a new quantum state $\ket{\chi'_0}_{n+1}$ and a new Grover operator $\mc{Q}'$, provided that $0 \leq f'(x) \leq 1$ for all $x$, which is required for the successful state preparation via Eq.~\eqref{eq:chi_0}.
Restricted by this constraint, the rescaling factor has to satisfy $0 \leq r f(x) \leq 1$ for all $x$.
To proceed further, we make the observation that the new Grover operator $\mc{Q}'$ implements a rotation by an angle $2 \theta'$ in the subspace $\mc{H}_{\chi'}$ spanned by good and bad renormalized states, as shown in Eq.~\eqref{eq:reflections}.
Under the commensurability condition
\begin{align}\label{eq:commesurate_theta}
 a' = \sin^2(\theta'),\quad \theta' = \pi l / k,\quad l \in \mb{Z},
\end{align}
applying the renormalized Grover operator $\mc{Q}'^k$ results in performing $l$ full rotations in $\mc{H}_{\chi'}$.
Such a commensurability condition can be achieved by fixing the renormalization factor as
\begin{align}\label{eq:r}
 r = a' / a = \sin^2(\theta') / a
\end{align}
where $\theta'$ is uniquely determined by the choice of the desired power $k$ and some integer number $l$.
Hence, we conclude that for a proper choice of $r$ satisfying $0 \leq r \leq 1 / \max\limits_x f(x)$, it is possible to rescale the function $f$ so that the $k$-th power of the corresponding Grover operator acts as identity in the subspace of good and bad states, \ie $\mc{Q}'^{k} = \mathbb{I}_{\chi'}$ in theory.

In practice, however, looking at Eq.~\eqref{eq:r} we see that finding the exact renormalization factor $r$ requires exact knowledge of the initial amplitude $a$ which, of course, we do not have.
However, as we show in the following, a loose estimate $a_{\trm{l}}$, obtained from a moderate number of MC samples of the initial state $\ket{\chi_0}_{n+1}$, is sufficient to get $\mc{Q}'^{k} \approx \mathbb{I}_{\chi'}$ and use it in adaptive VQAE.
Assuming that such a loose amplitude estimate is provided, a loose renormalization factor can then be expressed as $r_{\trm{l}} = a' / a_{\trm{l}}$, with $a'$ defined as in Eq.~\eqref{eq:commesurate_theta}.
Because of this imprecise estimation, the actual value of the amplitude after rescaling becomes $a'_{\trm{l}} = \sin^2 (\theta'_{\trm{l}}) = r_l a$ and the Grover operator performs $l$ full rotations only approximately, \ie the previous exact identity transforms into $\mc{Q}'^{k} \approx \mathbb{I}_{\chi'}$ with a typical phase error per Grover rotation of 
\begin{align}\label{eq:deviation}
 \delta \theta' =
 \theta' - \theta'_{\trm{l}} = 
 \theta' - \arcsin \sqrt{r_{\trm{l}} a} .
\end{align}
For an unbiased loose estimate with zero average, $\delta \theta'$ can be interpreted as a random error of zero mean.
After $k$ Grover rotations, this error becomes $k$ times as large.

Next, we use the VQAE algorithm to estimate the amplitude $a'_{\trm{l}}$ by means of the Grover operator $\mc{Q}'$ and the initial state $\ket{\chi'_0}_{n+1}$.
The variational approximation is performed at every $k$-th step, when the overlap
\begin{align}\label{eq:deviation_2}
 \Bra{{\chi'_0}}_{n+1} \Ket{{\chi'_k}}_{n+1} =
 \cos(2 k \theta'_{\trm{l}}) = \cos (2 k \delta \theta)
\end{align}
is expected to be the largest.
Here we use that $\theta'_{\trm{l}} + \delta \theta = \theta' = \pi l / k$.
Additionally, we assume that the PQC has the initial state $\ket{\phi_{\trm{init}}}_{n+1} = \ket{\chi'_0}_{n+1}$ so that the variational quantum state of Eq.~\eqref{eq:parametrized_circuit} reads
\begin{align}\label{eq:hybrid_circuit}
 \ket{\phi_{\trm{var}} (\vec{\lambda})}_{n+1} = \mc{U}_{\trm{var}}(\vec{\lambda}) \ket{\chi'_0}_{n+1} .
\end{align}
Having the PQC initialized to the identity at the beginning of each optimization step, the only role of the variational quantum circuit is to correct the deviation of Eq.~\eqref{eq:deviation} originating from an imprecise value of the renormalization constant $r_l$ and to bring the overlap of Eq.~\eqref{eq:deviation_2} as close to one as possible.
As a consequence, the variational optimization always starts from a good solution and therefore, in general, converges quicker to a better solution than na\"{i}ve VQAE.
This leads to a significant reduction in variational cost of adaptive VQAE compared to the na\"{i}ve version of the algorithm.

Finally, at the end of the calculation, a maximum likelihood estimation of $a'_{\trm{l}} = \sin^2(\theta'_{\trm{l}}) = r_{\trm{l}} a$ is obtained.
To go back to the original formulation of the problem and compare the results, we use the inverse transformation
\begin{align}
 \label{eq:scaling_inverse}
 \theta = \arcsin \sqrt{a},\quad a = a'_{\trm{l}} / r_l ,
\end{align}
where the renormalization constant $r_l$ has to be exactly the same as the one used for the function rescaling in order for the prior and posterior rescaling errors to cancel out.
This last step concludes the adaptive VQAE algorithm which is summarized in terms of pseudocode as Algorithm~\ref{algorithm_2}.

\begin{algorithm}[H]
\caption{Adaptive VQAE}
\label{algorithm_2}
\begin{algorithmic}
\Require functions $f$ and $p$, integer $k$
\State - Use $f$ to get a loose estimate $a_\trm{l}$
\State - Calculate $r_l = a' / a_{\trm{l}}$ and $f' = r_l f$
\State - Use $f'$ and $p$ to encode $\ket{\chi'_0}_{n+1}$ and $Q' = - \mc{R}_{\chi'} \mc{R}_{\trm{good}}$
\State - Use Algorithm~\ref{algorithm_1} to estimate the amplitude
\State $a'_{\trm{l}} = r_{\trm{l}} a$ associated with the state $\ket{\chi'_0}_{n+1}$
\State - Get the estimate for the original problem as $a = a'_{\trm{l}} / r_{\trm{l}}$
\end{algorithmic}
\end{algorithm}

We analyze the performance of the adaptive VQAE algorithm with a simplified variational ansatz consisting of only six single-qubit rotation gates and four CNOT gates, as shown in Fig.~\ref{fig:pqc} in dark blue color.
This simplified ansatz has only six parameters in total, which significantly reduces the number of variational queries as well as the effects of the noise due to finite sampling.
We determine the loose estimate of the amplitude $a_{\trm{l}}$ via $5 \times 10^5$ MC samples.
As a result, much smaller values of infidelity are achieved for $n_{f}$ being an order of magnitude smaller than in our na\"{i}ve VQAE computations.
We also note that for smaller values of $a$, the initial MC estimation of $a'$ gets worse and, as a consequence, more sweeps are required to ensure the convergence of the variational ansatz.

Our results for adaptive VQAE are presented in Fig.~\ref{fig:adaptive}, where we show the convergence of $\delta \theta$ as a function of $N_{\trm{q}}$ for $k = 10$.
The simulations use Adam with the initial learning rate $\beta = 10^{-3}$, $n_{f} = 100$, $n_{\trm{s}} = 100$, and $n_{\trm{p}} = 6$, resulting in $N_{\trm{var}/1} = 2 n_{f} n_{\trm{s}} n_{\trm{p}} = 1.2 \times 10^5$.
As in Fig.~\ref{fig:naive}, we compare with the classical MC scaling $\delta \theta \sim \mc{O}(N_{\trm{q}}^{-1/2})$ and the MLAE scaling $\delta \theta \sim \mc{O}(N_{\trm{q}}^{-3/4})$.
The major difference of the adaptive VQAE as compared to all previously studied methods is a large starting cost which corresponds to the amount of MC samples required for the evaluation of $a_{\trm{l}}$.
This starting cost, however, represents only an additive contribution to $N_{\trm{q}}$ and, hence, is insignificant in the regime of our interest when the number of queries gets large.
Additionally, we find that, thanks to a significant improvement of the number of query calls and the overall precision of the variational state, the resulting final error $\delta \theta$ of the adaptive VQAE algorithm surpasses the classical MC error.

Interestingly, we observe that in the regime of small $k$, the performance of the adaptive VQAE algorithm decreases.
This has several reasons.
Firstly, the precision of the maximum likelihood estimation decreases when the angle $\theta' = \pi l / k$ (where $l = 1$ in our case) becomes larger than $\pi / 4$, \ie for $k \leq 4$.
Hence, to perform an estimation with such small values of $k$, a different statistical inference technique has to be considered.
Secondly, for small values of $k$, the rescaling factor can become much larger than one and then leads to more efficient classical MC sampling.
Therefore, in calculations with $C = 0.1$ classical MC sampling performs better than adaptive VQAE for $k \leq 5$, corresponding to $r \gtrsim 7.508$.

\begin{figure}
\centering
\includegraphics[width=0.99\columnwidth]{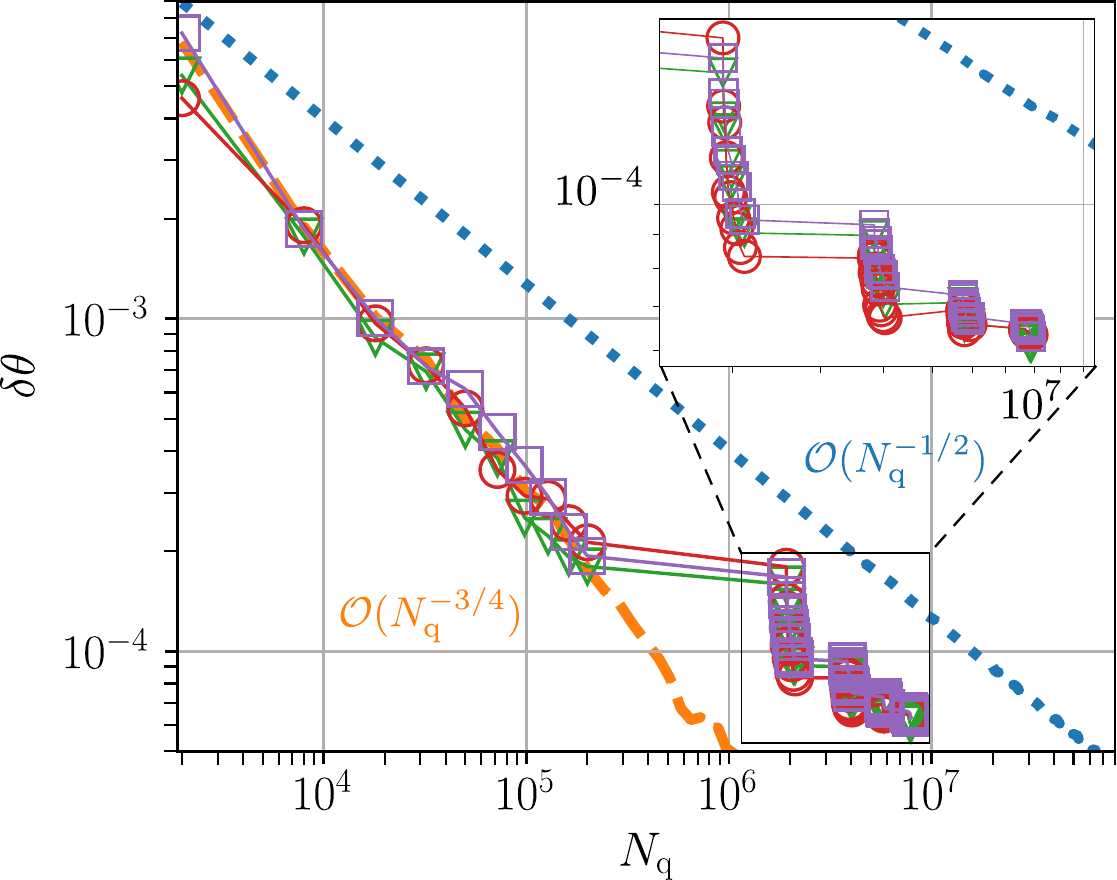}
\caption{\label{fig:adaptive2}
Amplitude estimation error $\delta \theta$ as a function of the number of queries $N_{\text{q}}$ for adaptive VQAE with $n = 8$ (green triangles), $10$ (red circles), and $12$ (purple squares) qubits for the Gaussian probability distribution in Eq.~\eqref{eq:problem_p_1}.
The remainder of the legend and the optimization procedure are the same as in Fig.~\ref{fig:adaptive}.
}
\end{figure}

To understand how adaptive VQAE performs for increasing qubit counts $n$, we have run the same simulations as in Fig.~\ref{fig:adaptive} for $n = 8$, $10$, and $12$.
The results for the Gaussian probability distribution are shown in Fig.~\ref{fig:adaptive2}.
For the shifted Cauchy-Lorentz and log-normal probability distributions we obtained results (not shown) lying on top of the ones in Fig.~\ref{fig:adaptive2}.
We find no significant dependence on $n$ in any of our results.
This is surprising as the cost function Eq.~\eqref{eq:objective_function} is global and therefore the vanishing gradient problem~\cite{CeEtAl20b} should lead to worse results for larger values of $n$.
We conjecture that the equally good performance of adaptive VQAE for all considered values of $n$ is due to the simple variational ansatz as well as the specific problems studied here.
Adaptive VQAE uses the simple ansatz shown in Fig.~\ref{fig:pqc} that consists of only 6 variational angles independent of $n$.
Additionally, the ansatz is composed of nearest-neighbour CNOTs and single-qubit $R_{y}$ rotation gates, i.e.\ not exact local 2-designs as in~\cite{CeEtAl20b}.
The problems studied here are expectation value calculations where an increased qubit count $n$ leads to an increased number of grid points $2^{n}$ for the discretized approximation, see Eq.~\eqref{eq:expectation_value}.
For the smooth functions considered here, we anticipate that the expectation value converges rapidly with increasing number of grid points.

\section{Discussion}
\label{sec:Discussion}

In this article, we provide numerical evidence that variational quantum algorithms based on constant-depth quantum circuits can be more efficient than classical MC sampling in the context of amplitude estimation.
The quantum circuits used for our numerical demonstrations, however, are still challenging for this generation of gate-based quantum computers.
Therefore, an exciting next step is to find other problems and applications for which VQAE has low quantum hardware requirements and can be realized on actual quantum devices.

We can imagine future applications for VQAE in several areas, including combinatorial optimization, quantum machine learning, and quantum chemistry.
In the context of combinatorial optimization, VQAE enables us to use constant-depth quantum circuits to carry out Grover search, which can find the optimal solution with a quadratic quantum speedup over brute-force search.
Here it is also enticing to study whether such a variational Grover search algorithm can benefit from filtering operators~\cite{AmEtAl21}.
In relation to quantum machine learning, VQAE has the potential to make it possible for current quantum devices to accelerate inference in Bayesian networks~\cite{LoYoCh14}, which can then be compared with state-of-the-art variational quantum algorithms for inference~\cite{BeEtAl21}.
With regards to quantum chemistry, the concept of VQAE can be combined with variational quantum phase estimation (VQPE)~\cite{PaMc19, StHuEv20, KlEtAl21} to realize VQPE with shallow circuits on actual quantum hardware.
In this context, one interesting application is to use accurate quantum chemistry results obtained with a quantum computer to train an ansatz for the exchange-correlation energy in density functional theory by means of machine learning~\cite{SnEtAl12, LuEtAl16, DiFe20}.

We anticipate that the efficiency of the VQAE algorithm can be further increased.
Firstly, it would be interesting to analyse whether a local cost function exists -- which can help mitigate the negative effect of barren plateaus~\cite{CeEtAl20b} -- that improves the variational optimization and reduces the required number of variational queries.
Secondly, the performance of our variational algorithm crucially depends on the maximum likelihood estimation procedure.
It would be interesting to investigate whether alternative approaches perform better, e.g.\ iterative QAE~\cite{GrinkoEtAl2019} or QoPrime AE~\cite{Giurgica-TironEtAl2020}.

\section*{Acknowledgements}

KP and ML are grateful to David Amaro and Marcello Benedetti for helpful discussions.

\bibliographystyle{unsrtnat}
\bibliography{bibliography.bib}

\end{document}